\documentclass[iop]{emulateapj}
\usepackage{amsmath}
\usepackage{amsfonts}
\usepackage{amssymb}
\usepackage{graphicx}
\usepackage{enumitem}
\begin{document}

\title{WASP-157b, a Transiting Hot Jupiter Observed with K2}
\author{T.~Mo\v{c}nik$^{1}$, D.~R.~Anderson$^{1}$, D.~J.~A.~Brown$^{2}$, A.~Collier~Cameron$^{3}$, L.~Delrez$^{4}$, M.~Gillon$^{4}$, C.~Hellier$^{1}$, E.~Jehin$^{4}$, M.~Lendl$^{5,6}$, P.~F.~L.~Maxted$^{1}$, M.~Neveu-VanMalle$^{6,7}$, F.~Pepe$^{6}$, D.~Pollacco$^{2}$, D.~Queloz$^{6,7}$, D.~S\'{e}gransan$^{6}$, B.~Smalley$^{1}$, J.~Southworth$^{1}$, A.~H.~M.~J.~Triaud$^{6,8}$, S.~Udry$^{6}$, and R.~G.~West$^{2}$}
\affil{$^{1}$Astrophysics Group, Keele University, Staffordshire, ST5 5BG, UK\\
$^{2}$Department of Physics, University of Warwick, Coventry CV4 7AL,
UK\\
$^{3}$SUPA, School of Physics and Astronomy, University of St Andrews, North Haugh, Fife, KY16 9SS, UK\\
$^{4}$Institut d'Astrophysique et de G\'{e}ophysique, Universit\'{e} de
Li\`{e}ge, All\'{e}e du 6 Ao\^{u}t, 17, Bat. B5C, Li\`{e}ge 1, Belgium\\
$^{5}$Space Research Institute, Austrian Academy of Sciences, Schmiedlstr. 6, 8042 Graz, Austria\\
$^{6}$Observatoire Astronomique de l'Universit\'{e} de Gen\`{e}ve
51 ch. des Maillettes, 1290 Sauverny, Switzerland\\
$^{7}$Cavendish Laboratory, J J Thomson Avenue, Cambridge, CB3 0HE,
UK\\
$^{8}$Institute of Astronomy, University of Cambridge, Cambridge, CB3 0HA, UK}
\email{t.mocnik@keele.ac.uk}
\shorttitle{Discovery of WASP-157b}
\shortauthors{Mo\v{c}nik et al.}

\begin{abstract}
We announce the discovery of the transiting hot Jupiter WASP-157b in a 3.95-d orbit around a $V = 12.9$ G2 main-sequence star. This moderately inflated planet has a Saturn-like density with a mass of $0.57 \pm 0.10$ M$_{\rm Jup}$ and a radius of $1.06 \pm 0.05$ R$_{\rm Jup}$. We do not detect any rotational or phase-curve modulations, nor the secondary eclipse, with conservative semi-amplitude upper limits of 250 and 20\thinspace ppm, respectively.
\end{abstract}


\section{INTRODUCTION}

The K2 mission \citep{Howell14} observes fields along the ecliptic. Some of those fields contain planets discovered by the WASP transit search \citep{Pollacco06} and so K2 is also observing some of the current WASP candidates. K2's high-precision photometry has allowed the detection of two additional close-in transiting exoplanets in the WASP-47 system \citep{Becker15} and the detection of starspot occultations in WASP-85 \citep{Mocnik16}, with several more WASP planets either recently observed or scheduled for future K2 campaigns.

The current planet formation and migration theories predict that hot Jupiters could not have formed \textit{in situ} and instead must have formed at larger distances from their host stars and migrated inwards after they were formed \citep{Lin96}. One of the tracers of hot Jupiters' migration processes is an obliquity, i.e. a misalignment angle between stellar rotation and planet's orbital axis. The aligned systems could result from a disk migration mechanism, whereas planet-planet scattering, Kozai mechanism and tidal dissipation are believed to play an additional role in shaping misaligned orbits of hot Jupiters (e.g. \citet{Triaud10}). Obliquity can be measured in two ways: a) spectroscopically by measuring the Rossiter-McLaughlin effect if the host star is bright enough \citep{Gaudi07}, and b) photometrically by tracing the recurring starspot occultation events if the host star is photometrically active \citep{Sanchis11}.

As of time of writing, the Transiting Extrasolar Planets Catalogue (TEPCat)\footnote{http://www.astro.keele.ac.uk/jkt/tepcat/tepcat.html} \citep{Southworth11} lists 95 planetary systems with known obliquities. A large portion of these, 75, are transiting hot Jupiters (defined here as planets with orbital periods less than 10 days and masses between 0.3 and 13\thinspace M$_{\rm Jup}$) orbiting stars brighter than $V = 13$. To try to fully understand how planets migrate it is crucial to expand the currently modest sample of planets with known obliquities by discovering and characterising additional nearby transiting hot Jupiters.

Many hot Jupiters exhibit radii well in excess of the predictions of the standard model of planets' cooling and contraction. This radius anomaly has been found to be negatively correlated with the planets' age \citep{Leconte09}. Several mechanisms have been proposed to account for the inflated radii, and are grouped into incident flux-driven mechanisms, tidal mechanisms and delayed contraction \citep{Weiss13}. However, none of the mechanisms has yet received a consensus. The radius anomaly currently remains unexplained and would benefit from extending a sample of known hot Jupiters.

Here we present the discovery of an inflated transiting hot Jupiter WASP-157b orbiting a fairly bright and photometrically inactive host star. In Section~2 we introduce the photometric and spectroscopic datasets. Basic properties of the host star are listed in Section~3, in Section~4 we discuss lightcurve modulations. Description of the spectrophotometric analysis and a list of system parameters are shown in Section~5, and a stellar age estimate is provided in Section~6.

\section{OBSERVATIONS}

WASP-157b was identified as an exoplanet candidate from observations with both WASP-South and SuperWASP-North over 2008--2010 (see Table~1). For a detailed description of the WASP telescopes, observing strategy, data reduction, and candidate identification and selection procedures, see \citet{Pollacco06,Pollacco08} and \citet{CollierCameron07a}. A transit was then observed with the TRAPPIST photometer \citep{Jehin11,Gillon11} on 2016 February 05.

WASP-157 (EPIC 212697709) was also observed with K2 in the long-cadence observing mode during Campaign 6 (from 2015 July 13 to 2015 September 30). Since the failure of the second out of four of \textit{Kepler}'s reaction wheels the spacecraft is no longer able to point as stably and the extracted photometry exhibits drift artefacts. \citet{Vanderburg14} presented a self-flat-fielding (SFF) correction procedure which corrects the drift artefacts by correlating the measured flux with the arclength of the spacecraft's drift. We retrieved the publicly available extracted and SFF-corrected lightcurve file provided by the K2 High Level Science Product K2SFF, accessible through Mikulski Archive for Space Telescopes (MAST). The applied SFF correction improved the median 6-hour photometric precision from 450 parts per million (ppm) to 26~ppm, which is comparable to original \textit{Kepler} precision for a similarly bright star. We normalized the downloaded K2SFF lightcurve using a {\scriptsize{KEPFLATTEN}} command which is part of the PyRAF tools for \textit{Kepler} (PyKE) version 2.6.2 \citep{Still12}. We used a low-order polynomial fit with a window and step size of 3 and 0.3 days, respectively. The reduced and normalized K2 lightcurve is shown in Figure~1.

\begin{figure*}
\includegraphics[width=\textwidth]{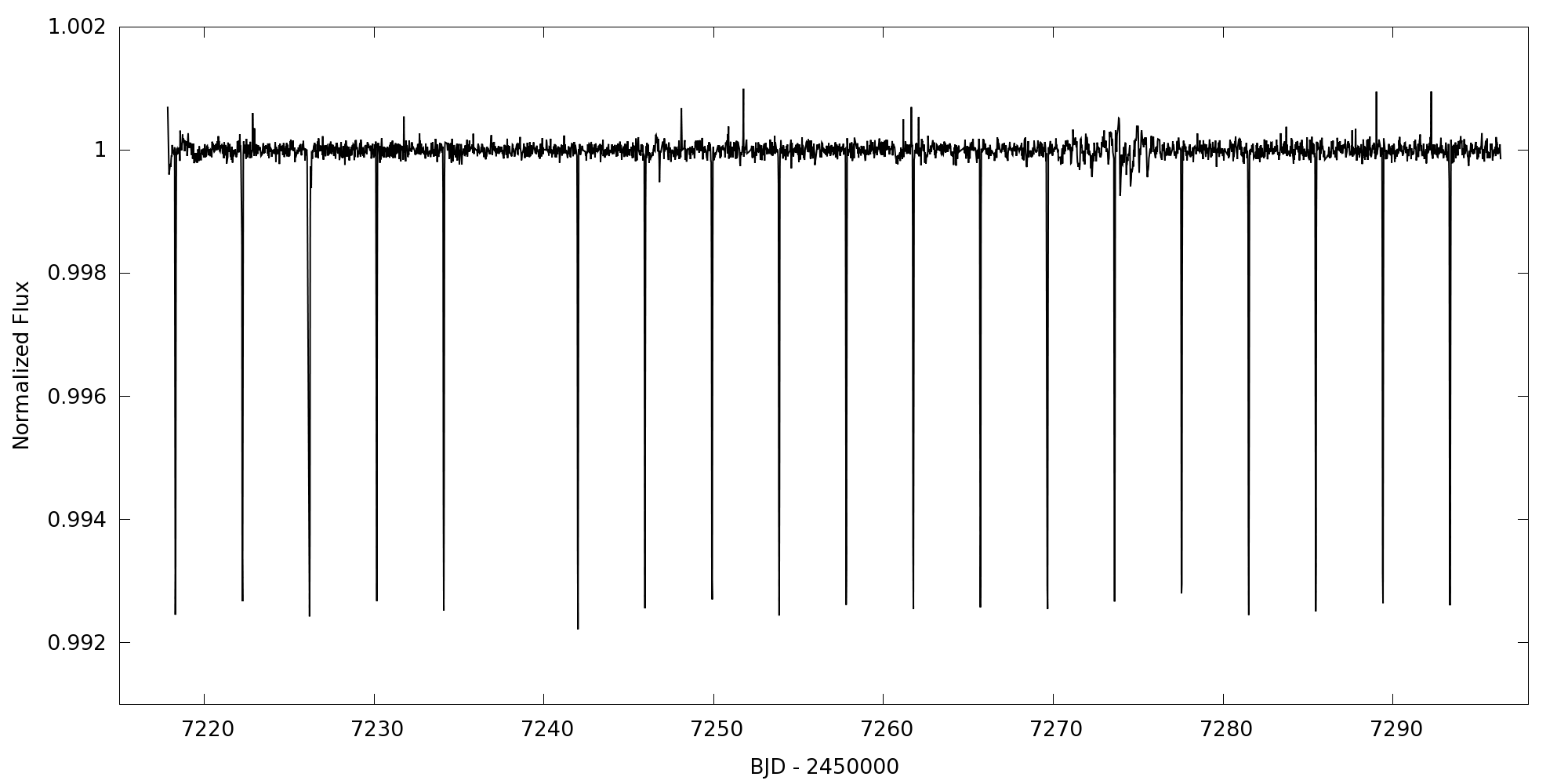}
\caption{Normalized K2 lightcurve of WASP-157. The observing Campaign 6 covers a total of 20 transits of WASP-157b. Note, however, that only 19 transits are visible because the data points obtained during the sixth transit were quality-flagged by the K2 preprocessing pipeline as a detector anomaly and were thus removed from the final lightcurve. The flux excursions in the final lightcurve are of instrumental origin, such as cosmic rays that have not been properly flagged by the preprocessing pipeline.}
\end{figure*}

Spectroscopic radial velocity (RV) follow-up was performed using the fibre-fed CORALIE spectrograph at the 1.2-m Swiss Euler Telescope at La~Silla \citep{Queloz00}. One spectrum was obtained in 2015 June and a further seven in 2016 February and March. We also obtained spectra on four consecutive nights in 2016 March with the HARPS spectrograph mounted at ESO 3.6-m telescope \citep{Mayor03}, also at La~Silla observatory (see Table~1). The spectroscopic data have been reduced with the standard HARPS and CORALIE reduction pipelines. The RVs were extracted with the weighted cross-correlation of each spectrum with a G2 mask and the simultaneous Th-Ar wavelength calibration reference (see \citet{Pepe02} for details). The resulting RVs and bisector spans (BS) are given in Table~2.

\begin{table}
\centering
\caption{Log of observations ($t_{\rm exp}$ is the effective exposure time, $N_{\rm data}$ the number of non-quality-flagged data points).}
\begin{tabular}{@{}clccc@{}}
\hline\hline
Instrument&Dates&Filter or&$t_{\rm exp}$ [s]&$N_{\rm data}$\\
&&wavelength&&\\
\hline
WASP&2008 May 04--&$R_{\rm c}$&30&42528\\
&2010 Jul 12&&&\\
K2&2015 Jul 13--&\textit{Kepler}&1625$^{a}$&3381\\
&2015 Sep 30&&&\\
TRAPPIST&2016 Feb 05&$I+z$'&15&529\\
CORALIE&2015 Jun 18--&390\thinspace --\thinspace 680\thinspace nm&1800&8\\
&2016 Mar 14&&&\\
HARPS&2016 Mar 01--&383\thinspace --\thinspace 693\thinspace nm&1000&4\\
&2016 Mar 04&&&\\
\hline
\end{tabular}
\begin{itemize}[leftmargin=0.35cm]
\setlength\itemsep{0cm}
\item[$^{a}$]Long-cadence K2 images are downlinked as a median of 270 frames of 6.02\thinspace s exposure and 0.52\thinspace s readout each, resulting in a final effective exposure time of 27.1\thinspace min obtained with a cadence of 29.4\thinspace min.
\end{itemize}
\end{table}

\begin{table}
\centering
\caption{CORALIE and HARPS radial velocities and bisector spans.}
\begin{tabular}{@{}lcr@{}}
\hline\hline
BJD $-$ 2450000&RV [km\thinspace s$^{-1}$]&BS [km\thinspace s$^{-1}$]\\
\hline
\noalign{\smallskip}
\textbf{CORALIE}\\
7191.64114&$-22.041 \pm 0.040$&$-0.073 \pm 0.079$\\
7430.78206&$-21.905 \pm 0.024$&$-0.033 \pm 0.048$\\
7431.74596&$-21.968 \pm 0.019$&$-0.060 \pm 0.038$\\
7432.85793&$-21.991 \pm 0.017$&$-0.048 \pm 0.034$\\
7433.78781&$-21.940 \pm 0.020$&$0.026 \pm 0.040$\\
7434.79450&$-21.877 \pm 0.020$&$-0.043 \pm 0.039$\\
7455.85032&$-22.016 \pm 0.020$&$-0.036 \pm 0.041$\\
7461.77389&$-21.909 \pm 0.014$&$-0.038 \pm 0.028$\\
\noalign{\smallskip}
\textbf{HARPS}\\
7448.73724&$-21.9940 \pm 0.0063$&$-0.019 \pm 0.012$\\
7449.89571&$-21.8974 \pm 0.0052$&$-0.030 \pm 0.010$\\
7450.70282&$-21.8708 \pm 0.0127$&$-0.002 \pm 0.025$\\
7451.78908&$-21.9657 \pm 0.0037$&$-0.018 \pm 0.007$\\
\hline
\end{tabular}
\end{table}

\section{SPECTRAL ANALYSIS}

We analysed the spectroscopic properties of the host star from a co-added HARPS spectrum with a final signal-to-noise ratio of 38. The analysis methods are described in \citet{Gillon09} and \citet{Doyle13}. We used the H$\alpha$ line to estimate the effective temperature ($T_{\rm eff}$), and the Na~{\sc i} D and Mg~{\sc i} b lines as diagnostics of the surface gravity ($\log g$). The projected rotational velocity ($v \sin I$) was determined by fitting the profiles of the Fe~{\sc i} lines after convolving with the HARPS instrumental resolution ($R$ = 120\thinspace 000) and a macroturbulent velocity adopted from the calibration of \citet{Doyle14}. The iron abundances were determined from equivalent width measurements of several clean and unblended Fe~{\sc i} lines and are given relative to the solar value presented in \citet{Asplund09}. The [Fe/H] abundance error includes that given by the uncertainties in $T_{\rm eff}$ and $\log g$, as well as the scatter due to measurement and atomic data uncertainties. We calculated the chromospheric activity index $\log R^{'}_{\rm {HK}}$ using the emission in the cores of the Ca~{\sc ii} H and K lines following \citet{Noyes84}. The resulting parameters are listed in Table~3. Only an upper limit is given for Li abundance since we do not detect the Li line at 6707.79\thinspace \AA, with an upper limit of 2\thinspace m\AA.

\begin{table}
\centering
\caption{Host star parameters derived from our spectroscopic analysis. The J2000 coordinates can be found in the 1SWASP identifier. \textit{V}-magnitude is taken from \citet{Hog00}.}
\begin{tabular}{@{}ll@{}}
\hline\hline
Parameter&Value\\
\hline
Identifiers&WASP-157\\
&1SWASP\thinspace J132637.24--081903.2\\
&2MASS\thinspace J13263727--0819033\\
&EPIC\thinspace 212697709\\
&TYC\thinspace 5544-596-1\\
Spectral type&G2V\\
\textit{V}&12.91\\
$T_{\rm eff}$&$5840 \pm 140$ K\\
$\log g$&$4.5 \pm 0.2$ cm\thinspace s$^{-2}$\\
$v \sin I$&$1.0 \pm 0.9$ km\thinspace s$^{-1}$\\
$v_{mac}$&$3.4 \pm 0.7$ km\thinspace s$^{-1}$\\
{[Fe/H]}&$+0.34 \pm 0.09$\\
$\log R^{'}_{\rm {HK}}$&$-4.8 \pm 0.2$\\
$\log \rm {A(Li)}$&$<$0.9\\
\hline
\end{tabular}
\end{table}

\section{LIGHTCURVE MODULATIONS}

We searched for any rotational modulation of WASP-157 using 2.2\thinspace yr of WASP lightcurves, following the procedure of \citet{Maxted11}. We found no modulations with semi-amplitudes above 2\thinspace mmag, which suggests that the host star is inactive. We also exclude any coherent rotational modulations above 250\thinspace ppm in the 79-d K2 fluxed lightcurve.

We also searched the phase-folded and binned K2 lightcurve for any phase-curve modulations including a secondary eclipse. For this task, we normalized the K2 lightcurve using larger values for the {\scriptsize{KEPFLATTEN}} window and step size of 5 and 1 day, respectively. These settings were a best compromise between minimizing the removal of the phase-curve modulations while still effectively removing the low-frequency incoherent lightcurve modulations. To check what portion of a potential phase-curve modulation gets removed using this normalization procedure, we injected various phase-curve modulation signals in the pre-normalized lightcurve, performed the normalization with the above mentioned settings, and found that we could recover injected signals from the normalized lightcurve at 75\% of their initial amplitudes. Based on $\chi^2$ statistics we could exclude phase-curve modulations with semi-amplitudes above 6\thinspace ppm at 95\% confidence level, after taking into account the 75\% throughput of the normalization procedure. However, the $\chi^2$-based upper limit does not account for the red noise, which is expected to be present in the binned phase-curve. Therefore, we provide a more conservative estimate of the phase-curve semi-amplitude upper limit of 20\thinspace ppm. Figure~2 shows the K2 lightcurve folded on the orbital period along with a reflection modulation model.

\begin{figure}
\includegraphics[width=8.5cm]{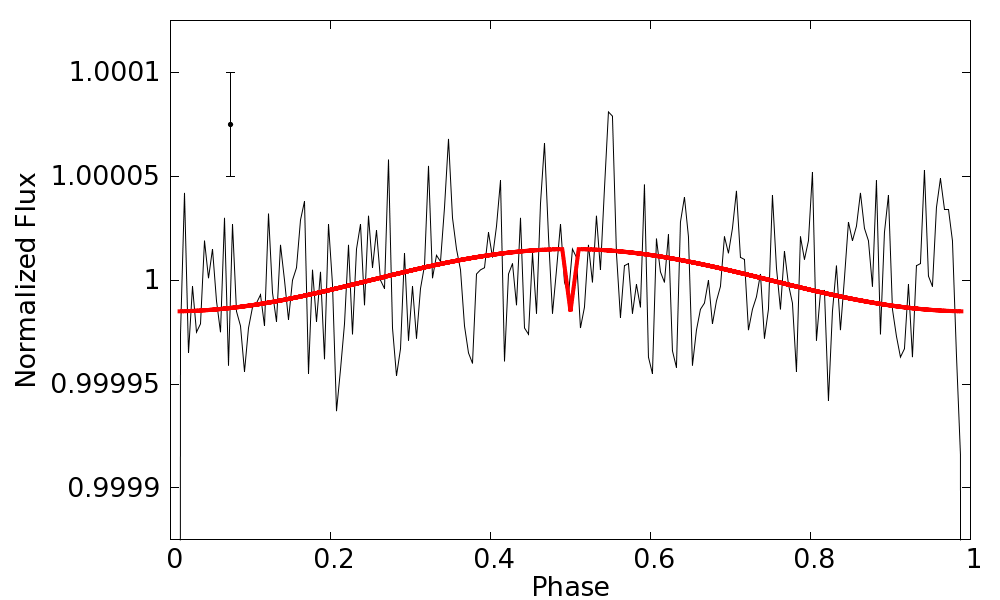}
\caption{Measured phase-curve of WASP-157, binned to 200 bins, (shown in black) and a simulated reflection phase-curve modulation with a semi-amplitude of 15\thinspace ppm (red). One representative error bar is shown on the left. The absence of any visible reflection modulation or secondary eclipse in the measured phase-curve of similar amplitude as shown in the simulated phase-curve suggests that the actual reflection semi-amplitude is lower than 20\thinspace ppm after taking into account the potential 25\% amplitude reduction due to the normalization procedure.}
\end{figure}

To look for any additional transiting planets we examined the periodogram obtained with PyKE tool {\scriptsize{KEPBLS}} which searches for periodic transits by utilizing the box least square fitting algorithm by \citet{Kovacs02}. After removing WASP-157b's transits from the K2 lightcurve by replacing the measured normalized flux values around transits with unity, we do not find any significant residual peaks in the period region between 0.5 and 30\thinspace d. The upper limit for additional transits is 250\thinspace ppm.

Transit-timing variations (TTVs) and transit-duration variations (TDVs) can also reveal additional planets in the system (e.g. \citet{Mazeh13} and references therein). However, due to the K2 long-cadence observing mode, TTV and TDV sinusoidal semi-amplitude upper limits are only weakly constrained to 1 and 5 minutes for periods up to 80 days, respectively.

\section{SYSTEM PARAMETERS}

To obtain stellar and planetary parameters we carried out a simultaneous analysis of the photometric and spectroscopic datasets. For analysing the transit shape we used primarily the K2 data, which have the best photometric precision, although with a 30-min cadence.

Since the effective exposure time of K2 long-cadence observations is significant compared to the time-scale at which the transit features occur, we performed the analysis where the transit model was computed with a sampling of 3\thinspace min, and integrated to the 30-min K2 sampling, to avoid any systematic errors on the transit parameters due to the ``smearing'' effect. The system parameters were fitted using Levenberg-Marquardt minimisation procedure and error bars computed with the Monte Carlo technique around the best-fit parameters. The analysis was performed with publicly available software {\scriptsize{JKTEBOP}}\footnote{http://www.astro.keele.ac.uk/jkt/codes/jktebop.html} \citep{Southworth04}. {\scriptsize{JKTEBOP}} is based on the earlier {\scriptsize{EBOP}} code by \citet{Popper81}, and has been modified, among many improvements and enhancements, to allow for a simultaneous fit of photometric data and RV measurements \citep{Southworth13}. The smearing correction can be seen by comparing the numerically integrated {\scriptsize{JKTEBOP}} fit (blue line) with the unintegrated fit (red line) in Figure~3. Quadratic limb-darkening was used in the {\scriptsize{JKTEBOP}} analysis.

Since {\scriptsize{JKTEBOP}} does not provide all the relevant system parameters, such as stellar and planetary mass, we produced a separate Markov chain Monte Carlo (MCMC) analysis using the code presented in \citet{CollierCameron07b} and further described in \citet{Pollacco08} and \citet{Anderson15}. We accounted for limb-darkening by interpolating within tables of four-parameter limb-darkening coefficients from \citet{Claret00,Claret04} and \citet{Sing10}, as appropriate for different filters used among the three photometric datasets (see Table~1). To avoid any bias because of the K2 smearing effect, we only included WASP and TRAPPIST photometry along with both spectroscopic datasets to obtain the remaining system parameters. We checked the compatibility between the {\scriptsize{JKTEBOP}} analysis using the K2 data, and the MCMC analysis of the ground-based observations alone, and found that the resulting parameters are fully consistent. For determining the epoch and period with the smallest uncertainty possible we ran another MCMC analysis using all the available datasets, which extended the photometric baseline from 79\thinspace d for K2 photometry to 7.8\thinspace yr and reduced the uncertainty on the period by a factor of 3.

For all the system parameters analysis we imposed a circular orbit since hot Jupiters are expected to circularise on a time-scale less than their age, and so adopting a circular orbit gives the most likely parameters (e.g. \citet{Anderson12}). To estimate the upper limit on eccentricity, we ran a separate MCMC analysis with eccentricity being fitted as a free parameter and checked the eccentricity distribution in the resulting MCMC chain.

We present the system parameters in Table~4 and superimpose the corresponding transit model on the measured lightcurves and RVs in Figure~3 and Figure~4, respectively.

The bottom panel of Figure~4 shows the measured bisector span (BS) for each spectrum, which characterises the asymmetry in the stellar line profiles \citep{Queloz01}. A significant correlation between BS and RV may indicate a transit mimic, such as a blended eclipsing binary \citep{Queloz01}. We, however, do not measure any statistically significant correlation between RVs and BSs.

\begin{table*}
\centering
\begin{minipage}{12cm}
\caption{System parameters for WASP-157b and its host star}
\begin{tabular}{@{}lccccc@{}}
\hline\hline
Parameter&Symbol&Value&Unit\\
\hline
Transit epoch$^{a}$&\textit{t}$_{\rm 0}$&2457257.803194 $\pm$ 0.000088&BJD\\
Orbital period$^{a}$&\textit{P}&3.9516205 $\pm$ 0.0000040&d\\
Area ratio&$(R_{\rm p}/R_{\star})^{2}$&0.00891 $\pm$ 0.00035&...\\
Transit width&\textit{t}$_{14}$&0.0811 $\pm 0.0043$&d\\
Impact parameter&\textit{b}&0.887$^{+0.015}_{-0.044}$&...\\
Orbital inclination&\textit{i}&84.93$^{+0.45}_{-0.21}$&$^{\circ}$\\
Orbital eccentricity&\textit{e}&0 (adopted; $<$0.11 at $2\sigma$)&...\\
Orbital separation&\textit{a}&0.0529 $\pm$ 0.0017&AU\\
Stellar effective temperature&\textit{T}$_{\rm eff}$&5838 $\pm$ 140&K\\
Stellar mass&\textit{M}$_{\star}$&1.26 $\pm$ 0.12&M$_\odot$\\
Stellar radius&\textit{R}$_{\star}$&1.134 $\pm$ 0.051&R$_\odot$\\
Stellar density&$\rho_{\star}$&0.86 $\pm$ 0.14&$\rho_\odot$\\
Planet equilibrium temperature$^{b}$&\textit{T}$_{\rm p}$&1339 $\pm$ 93&K\\
Planet mass&\textit{M}$_{\rm p}$&0.574 $\pm$ 0.093&M$_{\rm Jup}$\\
Planet radius&\textit{R}$_{\rm p}$&1.065 $\pm$ 0.047&R$_{\rm Jup}$\\
Planet density&$\rho_{\rm p}$&0.48 $\pm$ 0.10&$\rho_{\rm Jup}$\\
System RV&$\gamma$&--21.9522 $\pm$ 0.0026&km\thinspace s$^{-1}$\\
RV semi-amplitude&$K_1$&0.0616 $\pm$ 0.0038&km\thinspace s$^{-1}$\\
RV datasets offset&$\gamma_{\rm HAR}-\gamma_{\rm COR}$&0.0177 $\pm$ 0.0015&km\thinspace s$^{-1}$\\
\hline
\end{tabular}
\begin{itemize}[leftmargin=0.35cm]
\setlength\itemsep{0cm}
\item[$^{a}$]Epoch and period derived by fitting the photometric datasets from the K2 and all the available ground-based observations.
\item[$^{b}$]Planet equilibrium temperature is based on assumptions of zero Bond albedo and complete redistribution.
\end{itemize}
\end{minipage}
\end{table*}

\begin{figure}
\includegraphics[width=8.5cm]{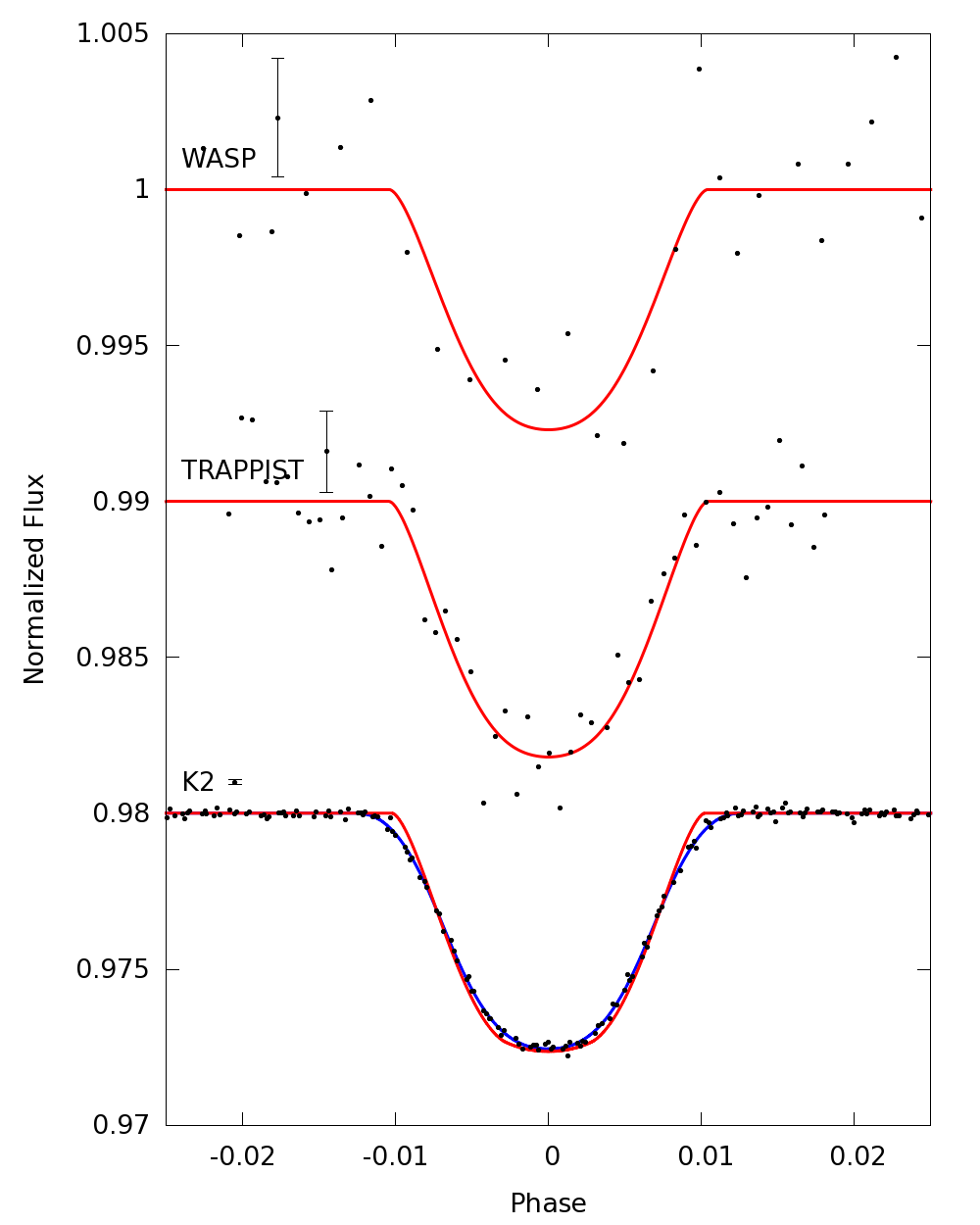}
\caption{Phase-folded lightcurves near the transit with best-fit transit models superimposed. The phase-folded WASP lightcurve and the TRAPPIST transit lightcurve have been binned by factors of 100 and 10, respectively. The K2 lightcurve was fitted using {\scriptsize{JKTEBOP}} with numerical integration (blue line) to correct for the smearing effect caused by the long effective exposure time. The corresponding {\scriptsize{JKTEBOP}} model without numerical integration is shown in red.}
\end{figure}

\begin{figure}
\includegraphics[width=8.5cm]{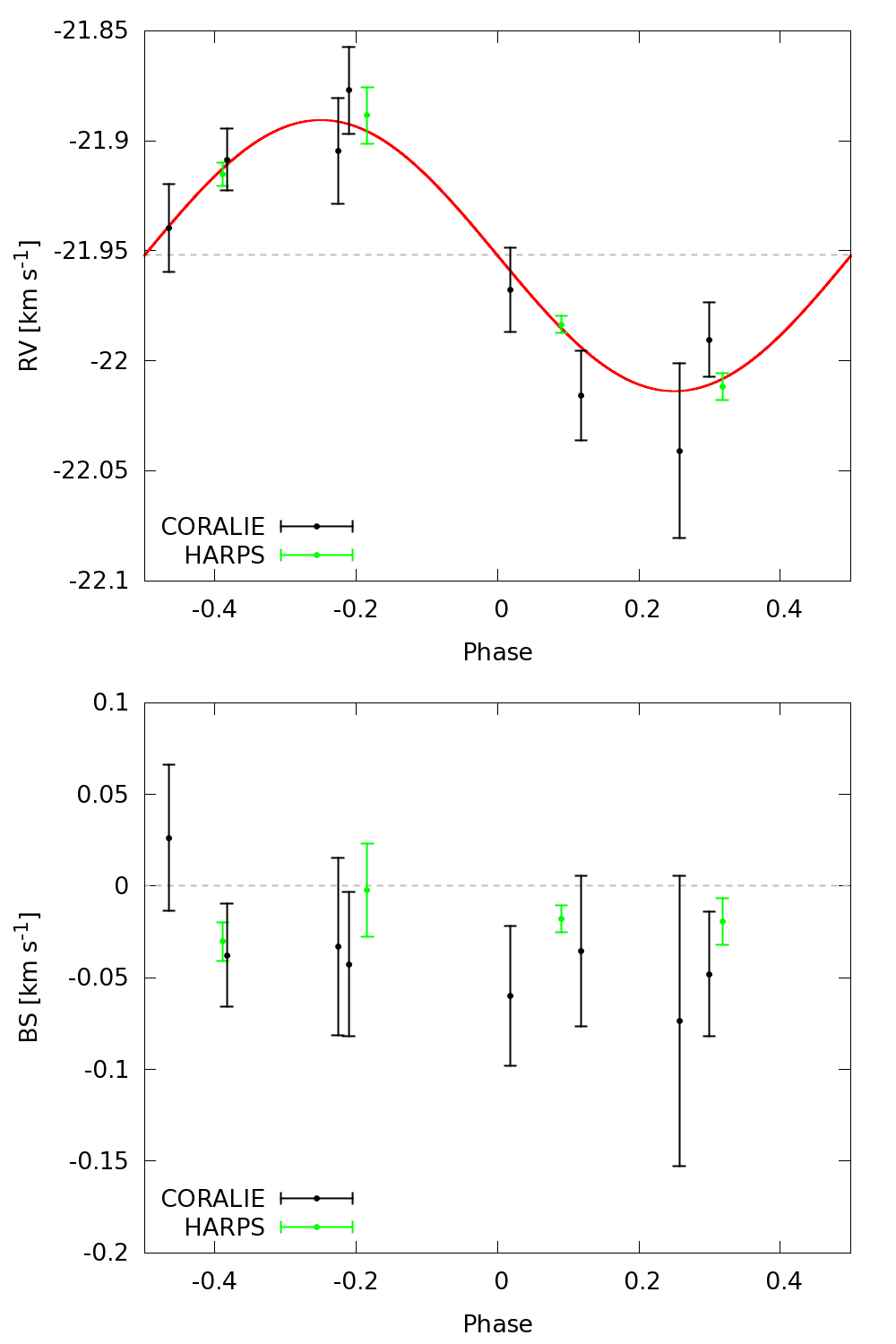}
\caption{Upper panel: Best-fit model of the CORALIE (black) and HARPS (green) RVs. Note that HARPS RVs have been offset by $-$0.0177\thinspace km\thinspace s$^{-1}$ (see Table~4). Bottom panel: CORALIE (black) and HARPS (green) bisector spans.}
\end{figure}

\section{STELLAR AGE}

An age constraint can be evaluated through a comparison to theoretical stellar models. As in \citet{Maxted16} we compare $\rho_{\star}$ and \textit{T}$_{\rm eff}$ to isochrones and evaluate the age of the star using the Bayesian mass and age estimator {\scriptsize{BAGEMASS}} by \citet{Maxted15}. The stellar evolution models used in {\scriptsize{BAGEMASS}} were calculated using the {\scriptsize{GARSTEC}} code \citep{Weiss08}. The best-fit stellar evolution track is shown in Figure~5 and provides an age estimate of $1.6^{+2.5}_{-0.8}$\thinspace Gyr.

Another age estimator is the chromospheric activity index $\log R^{'}_{\rm {HK}}$. Using the measured value from Table~3 and the relation between stellar age and activity for solar type stars from \citet{Mamajek08} we derive an age estimate of $3.2^{+3.4}_{-2.1}$\thinspace Gyr.

The measured Li abundance upper limit of $\log \rm {A(Li)} < 0.9$ (see Table~3) constrains the age of this Li-poor star only very weakly as being several Gyr old \citep{Sestitio05,Baumann10}.

\begin{figure}
\includegraphics[width=8.5cm]{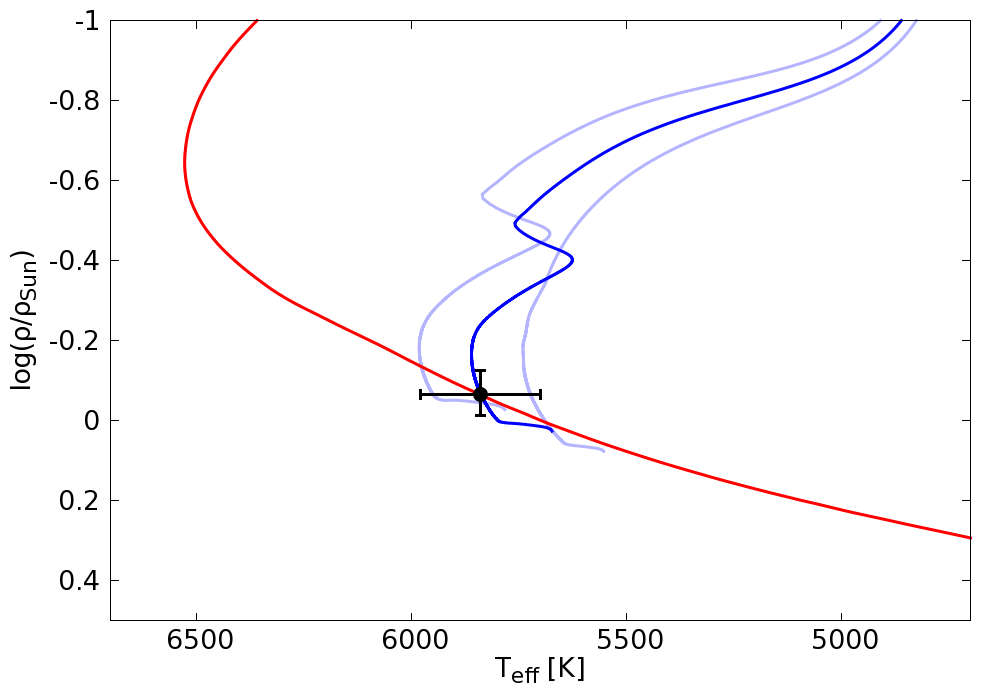}
\caption{WASP-157 host star in the $\rho_{\star}$ versus \textit{T}$_{\rm eff}$ plane compared to the best-fit evolution track (dark blue line) and isochrone of 1.6\thinspace Gyr (red line). Two light blue lines  correspond to stellar models at 5 percent higher and lower mass for comparison.}
\end{figure}

\section{CONCLUSIONS}
WASP-157b is very much a typical hot Jupiter. The orbital period of 3.95\thinspace d is typical, while the moderately inflated size (1.1\thinspace R$_{\rm Jup}$ for 0.57\thinspace M$_{\rm Jup}$) is also characteristic of hot Jupiters. The impact parameter is higher than average, leading to a shorter, v-shaped transit. WASP-157 is notable for now being one of fewer than 20 hot-Jupiter hosts with $V < 13$ to have a \textit{Kepler}-quality lightcurve.

Our measured and derived stellar parameters from Tables~3 and 4 agree within about 1$\sigma$ with the values provided by the K2 Ecliptic Planet Input Catalogue (EPIC) \citep{Huber16}. The only discrepant parameter is metallicity, which has only been estimated statistically by the EPIC because no spectroscopic input was provided for this star.

\acknowledgements{We would like to thank the anonymous referee for their comments which led to improving this paper. We gratefully acknowledge the financial support from the Science and Technology Facilities Council (STFC), under grants ST/J001384/1, ST/M001040/1 and ST/M50354X/1. WASP-South is hosted by the South African Astronomical Observatory and SuperWASP-North by the Isaac Newton Group of Telescopes and the Instituto de Astrofisica de Canarias. We are grateful for their ongoing support and assistance. TRAPPIST is funded by the Belgian Fund for Scientific Research (Fond National de la Recherche Scientifique, FNRS) under the grant FRFC 2.5.594.09.F, with the participation of the Swiss National Science Fundation (SNF). The Euler Telescope is operated by the University of Geneva thanks to a grant by the Swiss National Science Foundation. LD acknowledges support of the F.R.I.A. fund of the F.R.S-FNRS. MG is Research Associate at the Belgian F.R.S-FNRS. This paper includes data collected by the K2 mission. Funding for the K2 mission is provided by the NASA Science Mission directorate. This work made use of PyKE \citep{Still12}, a software package for the reduction and analysis of \textit{Kepler} data. This open source software project is developed and distributed by the NASA Kepler Guest Observer Office.}

\bibliographystyle{apj}
\bibliography{bibliography}

\end{document}